# A set of R packages to estimate population counts from mobile phone data

Bogdan OANCEA[1]
David SALGADO[2]
Luis SANGUIAO SANDE[3]
Sandra BARRAGAN[4]

**Abstract**

In this paper, we describe the software implementation of the methodological framework designed to incorporate mobile phone data into the current production chain of official statistics during the ESSnet Big Data II project. We present an overview of the architecture of the software stack, its components, the interfaces between them, and show how they can be used. Our software implementation consists in four R packages: **destim** for estimation of the spatial distribution of the mobile devices, **deduplication** for classification of the devices as being in 1:1 or 2:1 correspondence with its owner, **aggregation** for estimation of the number of individuals detected by the network starting from the geolocation probabilities and the duplicity probabilities and **inference** which combines the number of individuals provided by the previous package with other information like the population counts from an official register and the mobile operator penetration rates to provide an estimation of the target population counts.

## 1. Introduction

Mobile network data represent a new and powerful source of information for the production of official statistics in several fields: population density, population mobility, tourism, transportation etc. The European Statistical System is under way to build a general framework to incorporate the mobile phone data into the current production of official statistics (Ricciato, 2018). As a part of the ESSnet Big Data II project, Work Package I (https://ec.europa.eu/eurostat/cros/content/WPI_Mobile_networks_data_en), we have made a proposal that constitutes the first step towards the construction of the so-called ESS Reference Methodological Framework for Mobile Network Data (Ricciato, 2018). Our methodological proposal is described in Salgado et al. (2020) and it is an end-to-end statistical production process. Shortly speaking, our methodological approach starts from a geographical territory represented by a map, a telecommunication network configuration, a population of individuals carrying 0, 1, or 2 mobile devices during their movement, and a reference grid with rectangular tiles overlapped on the map and provides an estimation of the population counts using the network event data (the data generated by the communication between mobile devices and the base transceiver stations - BTS) and some auxiliary data.

---

[1] National Institute of Statistics Romania and University of Bucharest, Romania email: bogdan.oancea@insse.ro, bogdan.oancea@faa.unibuc.ro – corresponding author
[2] National Institute of Statistics, Spain, Madrid, email: david.salgado.fernandez@ine.es
[3] National Institute of Statistics, Spain, Madrid, email: luis.sanguiao.sande@ine.es
[4] National Institute of Statistics, Spain, Madrid, email: sandra.barragan.andres@ine.es

Since functional modularity in the statistical process is a central element when working with new data sources that are technology dependent, our methodological approach is organized based on the modularity and abstraction principles. This means the division of the production process into separate parts/modules, which are designed so that their interaction takes place only through the interfaces, making the internals of each module independent of the rest of modules.

Organizing the modules in such a way to minimize their interaction is achieved through layering/stacking them in a hierarchy. Following this idea, the software implementation uses a layered design, which is one of the most common architecture patterns in software design (Richards, 2015). Besides closely following the architecture of the methodological framework, the layered design has other advantages too: it is easy to develop, maintain and test, and changing the implementation of one layer does not affect the rest of the components. The layered structure of the software implementation is shown in figure 1. Our stack of software modules is organized as follows.

**Data acquisition and preprocessing layer**. This layer deals with capturing the network events and applying a series of preprocessing operations to bring them into a form that can be statistically exploited. It is strongly dependent on the mobile network technology, which can vary among different MNOs and geographical regions. This component is not implemented in our software stack because we lacked access to a real mobile network during the process of writing the code. Instead, we used a mobile network data simulator that is described in detail by Oancea et al. (2019).

**Geolocation layer**. The main purpose of this module is to exploit the network event data and to derive the probability of localization for each device at the level of geographical units. This is performed using a Hidden Markov Model and is implemented in the R package **destim** available at the following URL: https://github.com/MobilePhoneESSnetBigData/destim. It provides the location probability for each device as well as the joint location probabilities.

**Deduplication layer**. The purpose of this component is to classify each device *d* as corresponding to an individual with only one device (1:1) or to an individual with two devices (2:1). We used a probabilistic classification, thus assigning a probability $p_d^{(n)}$ that a device *d* is carried by an individual with *n* devices. We only allowed a maximum number of 2 devices per individual, but the generalization is easy. The layer is implemented in the R package **deduplication** available at the following URL: https://github.com/MobilePhoneESSnetBigData/deduplication.

**Aggregation layer**. The function of this layer is to estimate the number of detected individuals starting from the location probabilities and making use of the duplicity probability for each device provided by the previous layers. It is implemented in the R package **aggregation** available at the following address: https://github.com/MobilePhoneESSnetBigData/aggregation.

**Inference layer**. The role of this layer is to compute the probability distribution for the number of individuals in the target population conditioned on the number of individuals detected by the network and some auxiliary information coming from mobile network operator penetration rates and population registers. It is implemented in the R package **inference** available at the following address: https://github.com/MobilePhoneESSnetBigData/inference.

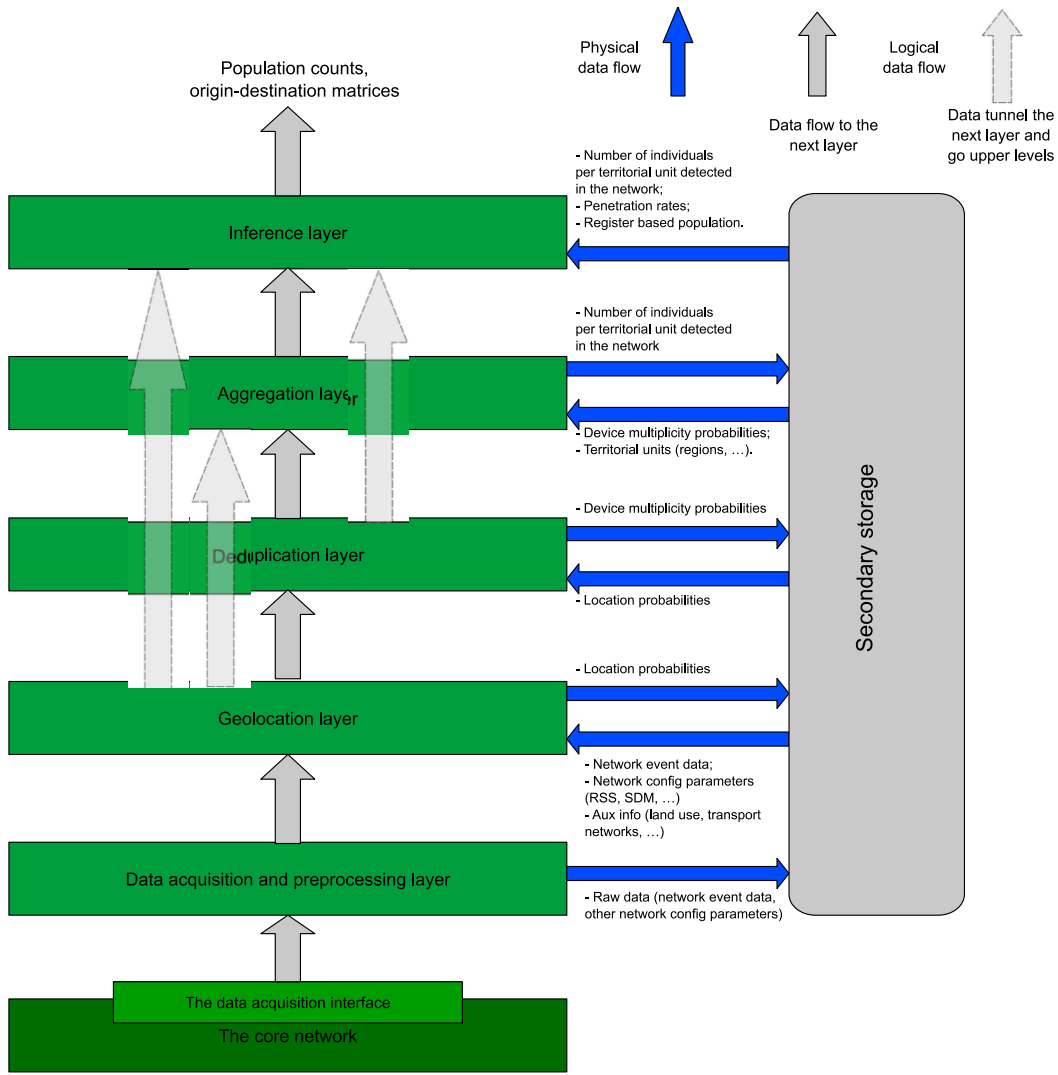

Figure 1. The layered structure of the software implementation.

The rest of the paper is organized as follows. In section 2 we provide details about the **destim**, **deduplication**, **aggregation** and **inference** R packages, in section 3 we discuss the computational efficiency and the scalability of these R packages and section 4 concludes our paper.

## 2. The R packages

In this section we provide a short description of each layer (implemented in an R package) in terms of input and output data and give examples how to use each package. The theoretical background of the methods implemented in these R packages is described in Salgado et al. (2020) and an interested reader should first consult this work.

All the modules of the software stack are entirely decoupled. One layer receives some input data from the layer(s) below it and provides data to the upper layer(s) as output. Setting a clear format for these data sets that flow from one layer to another make the layers independent and easy to change their implementation. The single request is to adhere to the format of the datasets passed as input and to the ones provided as the output. We define the interface between consecutive layers as the format of the datasets flowing from one layer to another. For the initial versions of our R packages the structure of the csv files are fixed but there is an ongoing work to standardize the formats (column names, data types, accepted values, etc.) and define these standards using schema definition (xsd) and xml files. Then, all R packages from our software stack will first read the files describing the structure of the data and after that read the files themselves and process the data.

We mention that besides these data that flow vertically, there are other general parameters available to all layers from the secondary storage. We do not deal here with the input data for the first layer (Data acquisition) since it is technology dependent and out of the control of statisticians.

## 2.1 destim

**destim** package (Sanguiao et al., 2020) exploits the network event data and derive localization probability for each device at the level of geographical units using a Hidden Markov Model (HMM). The estimation of the localization probabilities requires first model construction and then model fitting. The inputs and outputs of this R package are:
- *Inputs*:
    - Network events (*a csv file*);
    - Signal strength/signal dominance needed to compute the emission probabilities for HMM (*a csv file*);
    - Some general parameters: grid/tile size, the sequence of time instants;
- *Outputs* *(as csv files)*:
    - Location probabilities for each device, each tile and time instant – a file per device;
    - Joint location probabilities for each device, each tile and consecutive time instants.

The network events file is a csv file with the following structure:

```
t, Antenna ID, Event Code, Device ID
```

where `t` is the time instant when the event was generated, `Antenna ID` a unique ID of the antenna[5] (BTS) that recorded this event, `Event Code` an integer that represents the code of the network event and `Device ID` a unique ID of the device that generated the event.

The signal strength or the signal quality/dominance (Salgado et al. 2020, Oancea et al. 2019) is a key information used to compute the location likelihood and it must be computed in the center of each tile of the grid. The values in this file depend on the technical parameters of the network as well as on the characteristics of the geographical region (open field, cities etc.). The file providing these values has the following structure:

```
Antenna ID, Tile0, Tile1, Tile2,..., TileN-1
```

where the `Antenna ID` is the unique ID of the antenna and the rest of the columns represents the IDs of the tiles of the grid.

The first output of the **destim** package is mainly a matrix with *N* rows and *nTimes* columns where *N* is the total number of tiles in the grid and *nTimes* is the total number of time instants when the network events were recorded. A value *p(i, j)* in this matrix is the location probability for tile *i* and time instant *j* for the device under consideration. Since this is a sparse matrix, we used a format similar to the Coordinate Text File (Boisvert et al. 1997) to store these matrices. The structure of this output file is simple:

```
time, tile, probL
```

where `time` is the time instant, `tile` is the tile ID and `probL` is the posterior location probabilities (only non zero probabilities are stored).

---

[5] We use loosely the term antenna in this paper as a radio cell. A given BTS (antenna) can provide coverage for multiple radio cells over the territory depending on its configuration. The software provides event information for each radio cell so that if a given BTS (antenna) has three radio cells, the dataset will have entries for each radio cell of this same antenna.

The joint location probabilities files stores the probability of being in tile *i* at time instant *t - 1* and tile *j* at time instant *t* for all combinations of consecutive time instants and tiles. We store again only the non-zero probabilities. The structure of these files is:

```
time_from, time_to, tile_from, tile_to, probL
```

where `time_from` and `time_to` are the initial and final time instant of a transition, `tile_from` and `tile_to` the initial and destination tile ID of a transition and `probL` the probability that the device under consideration is located in `tile_from` at `time_from` and in `tile_to` at `time_to`.

Producing location probabilities with **destim** requires a sequence of steps:
- Read all input files;
- Compute the event location probabilities;
- Prepare the initial state distribution of the HMM;
- Compute the posterior location probabilities and joint location probabilities by fitting an HMM for each device.

**destim** provides complete examples how to produce the location probabilities starting from the input data. Because the example scripts are rather long, we will not reproduce them here but invite interested users to consult them after installing the package.

## 2.2 deduplication

**deduplication** package (Oancea et al., 2020a) computes a device multiplicity probability $p_d^{(n)}$ for each mobile device *d*, i.e. the probability that a device *d* is carried by an individual carrying *n* devices. Currently *n* could be 0, 1, 2. This package implements two approaches to compute the duplicity probabilities: a Bayesian approach based on network events (with 2 variants: *1-to-1* and *pairs*) and a *trajectory* similarity approach. A detailed theoretical description of these methods can be found in (Salgado et al. 2020).

The inputs of the **deduplication** package are the outputs of the previous layer, i.e. the two sets of files with the location probabilities and joint location probabilities for each device. Besides these two data sets, the **deduplication** package also needs some files that come directly from the data acquisition and preprocessing layer: the network events file and the antennas cell file. The network event file structure was already presented, and the antennas cell file has the following structure:

```
AntennaId,Cell Coordinates
```

where `Cell Coordintates` is a WKT string describing the outer boundaries of the cell.

An additional input of this layer is an information from an external source that gives the a priori probability of a person to hold two mobile devices.

The output of the **deduplication** package is a single csv file that contains the duplicity probability for each device. It has two columns:

$$\texttt{deviceID, dupP}$$

where `deviceID` is the unique ID of a device and `dupP` is the probability of the device to be in a 2-1 correspondence with its owner. There is one row for each device registered by the network.

Arriving at the final duplicity probability table involves several intermediate steps. In order to hide all these details from the user the **deduplication** package provide the function *computeDuplicity*() that is easy to use. Below we show a full example using this function. All the input files used in this example are provided with the **deduplication** package.

Firstly, we set the folder where the necessary input files are stored:

```
path_root <- 'extdata'
```

Next, we set the grid file name, i.e. the file where the grid parameters are found:

```
gridfile <- system.file(path_root, 'grid.csv', package = 'deduplication')
```

Then we set the event file name, i.e. the file with network events:

```
eventsfile <- system.file(path_root, 'AntennaInfo_MNO_MNO1.csv', package = 'deduplication')
```

We also need to set the file name where the signal strength/dominance for each tile in the grid is stored:

```
signalfile <- system.file(path_root, 'SignalMeasure_MNO1.csv', package = 'deduplication')
```

The antenna cells file is needed to build the list of neighboring antennas. This file is needed only if the duplicity probabilities are computed using *pairs* or *trajectory* methods:

```
antennacellsfile <- system.file(path_root, 'AntennaCells_MNO1.csv', package = 'deduplication')
```

Finally, we set the file with the simulation parameters used to produce the data set:

```
simulationfile <- system.file(path_root, 'simulation.xml', package = 'deduplication')
```

Now we can compute the duplicity probabilities using one of the three methods:

Using the *1to1* method:

```
out1 <- computeDuplicity(method = "1to1", gridFileName = gridfile, eventsFileName = eventsfile, signalFileName = signalfile, simulatedData = TRUE, simulationFileName = simulationfile)
```

Using the *1to1* method with *lambda* parameter (note that the value given here is for convenience):

```
out1p <- computeDuplicity(method = "1to1", gridFileName = gridfile, eventsFileName = eventsfile, signalFileName = signalfile, simulatedData = TRUE, simulationFileName = simulationfile, lambda = 0.67)
```

Using the *pairs* method:

```
out2 <- computeDuplicity(metho = "pairs", gridFileName = gridfile,
eventsFileName = eventsfile, signalFileName = signalfile, antennaCellsFileName
= antennacellsfile, simulationFileName = simulationfile)
```

Using the *trajectory* method:

```
out3 <- computeDuplicity(method = "trajectory", gridFileName = gridfile,
eventsFileName = eventsfile, signalFileName = signalfile,antennaCellsFileName
= antennacellsfile, simulationFileName = simulationfile, path= path_root)
```

## 2.3 aggregation

The aggregation process connects the number of devices detected by the network with the number of individuals in the network. There are two main goals in this process:
    a. The number of detected individuals;
    b. The origin - destination matrix.

For this purpose, the aggregation process uses a probabilistic approach to carry forward the uncertainty already present in the preceding stages all along the end-to-end process. Both the geolocation of network events and the device' duplicity classification are probabilistic in nature, therefore, a priori it is impossible to provide a certain number of individuals in a given territorial unit.

For this reason, the methodology behind this process focuses on the probability distribution of the number of individuals detected by the network. Having a probability distribution amounts to having all statistical information about the random phenomenon and one can choose any point estimation (mean, median, mode) together with uncertainty measures (coefficient of variation, standard deviation, credible intervals). The random variable modeling the number of detected individuals is Poisson multinomial distributed as it is shown in Salgado et al. (2020). The properties and software implementation of this distribution are not trivial, and we used a Monte Carlo simulation method by convolution to generate random variates according to this distribution.

The **aggregation** package (Oancea et al. 2020b) provides a single function to generate random variates according to the Poisson multinomial distribution: *rNnetEvent*(). Thus, all the details about intermediate computations are hidden from the users. The input data needed by this function are:
- n: the number of the random values to be generated;
- dupFileName: the file with the duplicity probabilities for each device;
- regsFileName: the file defining the geographical regions where we intend to aggregate the number of individuals;
- postLocPath: the path to the directory where the files with the posterior location probabilities for each device are found;
- prefix: the name prefix of these files (which are the output of the **destim** package);
- times: a vector with the values of time instants.

We provide a complete set of files with all the input data in the *extdata* folder of this package. The raw data used to produce these files are given by our simulation software and by **destim** and **deduplication** packages. Below we show a simple example of how to use the *rNnetEvent()*.

```
# set the folder where the necessary input files are stored
path    <- 'extdata'
prefix <- 'postLocDevice'

# get the series of time instants from the simulation.xml file.
simParamsFileName <- system.file(path, 'simulation.xml', package =
'aggregation')
simParams <- deduplication::readSimulationParams(simParamsFileName)

time_from <- simParams$start_time
time_to   <- simParams$end_time
time_incr <- simParams$time_increment
times     <- seq(from = time_from, to = time_to-time_incr, by = time_incr)

# set the duplicity probabilities file name,
#i.e. the file with duplicity probability for each device
dpFile <- system.file(path, 'duplicity.csv', package = 'aggregation')

# set the regions file name, i.e. the file defining the regions for which
# we need the estimation of the number of individuals detected by network.
rgFile <- system.file(path, 'regions.csv', package = 'aggregation')

# set the path to the posterior location probabilities files
pathLoc <- system.file(path, package = 'aggregation')

# set the number of random values to be generated
n <- 1e3
# call rNnetEvent
nNet <- rNnetEvent(n = n, dupFileName = dpFile, regsFileName = rgFile,
postLocPath = pathLoc, prefix = prefix,times = times)
head(nNet)
time region N iter
1: 1 1 11 1
2: 1 1 11 2
3: 1 1 10 3
4: 1 1 13 4
...
```

In the `nNet` output dataframe, `time` represents the time instant, `region` is the region number and `N` is the number of individuals. For each combination of time instant and region there are several random values generated for `N` and their index is given in `iter` column.

The construction of the probability distribution for the number of individuals detected by the network can be easily generalized to the number of individuals detected by the network moving between territorial units. The **aggregation** package provides a single function to generate random variates that can be used to compute an estimation of the number of individuals moving from one region to another. The input parameters of this function are:
- n: the number of the random values to be generated;
- dupFileName: the file with the duplicity probabilities for each device;
- regsFileName: the file defining the geographical regions where we intend to aggregate the number of individuals;
- postLocJointPath: the path to the directory where the files with the posterior joint location probabilities for each device are found;

- prefix: the name prefix of these files.

The files containing the joint probabilities are also a result of the **destim** package.

```
# set the folder where the necessary input files are stored
path             <- 'extdata'
postLocJointPath <- system.file(path, package = 'aggregation')
prefixJ          <- 'postLocJointProbDevice'

# set the duplicity probabilities file name,
# i.e. the file with duplicity probability for each device
dpFile <- system.file(path, 'duplicity.csv', package = 'aggregation')

# set the regions file name, i.e. the file defining the regions for which
# we need the estimation of the number of individuals detected by network.
rgFile <- system.file(path, 'regions.csv', package = 'aggregation')

# generate n random values
n      <- 1e3
nnetOD <- rNnetEventOD(n = n, dupFileName = dpFile, regsFileName = rgFile,
postLocJointPath = postLocJointPath, prefix = prefixJ)
head(nnetOD)

time_from time_to region_from region_to Nnet iter
1: 0 10 1 1 18.0 1
2: 0 10 1 1 18.5 2
3: 0 10 1 1 19.0 3
...
```

The number of individuals moving from a region to another is given in the `nnetOD` dataframe and the name of the columns are self-explanatory. For each distinct combination (`time_from, time_to, region_from, region_to`) there are several random values generated for the number of individuals and their index is given in `iter` column.

## 2.4 inference

The inference process connects the number of individuals in the network with the number of individuals in the target population. The main goal of the **inference** package (Oancea et al. 2020c) is the computation of the probability distribution for the number of individuals in the target population conditioned on the number of individuals detected by the network and some auxiliary information which is necessary to provide a meaningful inference on the target population. The auxiliary information is provided by the penetration rates of MNOs (ratio of number of devices to number of individuals in the target population) and the register-based population at each region $r$. There are three main functionalities provided by the inference package:
- Population at the initial time $t_0$.
- The dynamical approach i.e. population at $t > t_0$.
- The Origin-Destination matrices.

The inputs of this package are:

- the posterior location probabilities, one file per device (provided by **destim**);
- a csv file with the duplicity probabilities for each device, (provided by **deduplication**);
- a csv file defining the geographical regions;
- a csv file with the general parameters defining the grid;
- a csv file with information from a population register, giving the population count for all geographical regions under consideration at the considered initial time;
- a csv file with the penetration rate of the mobile network operator for all geographical regions under consideration;
- a csv file with the number of individuals for each time instant and region which is an output from the **aggregation** package;
- a csv file with the number of individuals moving from a region to another which is also an output from the **aggregation** package.

The information from the population register is organized on two columns as showed below:

```
region, N0
1, 38
2, 55
3, 65
...
```

Here, `region` is the region number and `N0` is the population count in the corresponding region. The penetration rate file has a similar structure:

```
region, pntRate
1, 0.3684211
2, 0.4
3, 0.4153846
...
```

The second column, `pntRate`, is the penetration rate obtained from the MNO.

The population counts can be modelled with three probability distributions: Beta Negative Binomial distribution, Negative Binomial distribution and the State Process[6] Negative Binomial distribution (for a detailed demonstration see Salgado et al. 2020).

For the population at the initial time instant $t_0$ the **inference** package generates two tables: one with descriptive statistics about the distribution and another one with the random values of the population counts generated for each region. The descriptive statistics of the population count distribution are showed below:

```
region Mean Mode Median Min Max Q1 Q3 IQR SD CV CI_LOW CI_HIGH
1 43 33 39 11 133 30 51 21 17.90 41.96 21.00 73.00
2 59 51 56 21 126 47 68 21 15.85 27.09 37.00 88.00
3 82 68 78 38 185 68 94 26 20.84 25.38 54.97 121.00
...
```

The random values generated according to the corresponding distribution are organized in a table with the following structure:

---

[6] Also a Negative Binomial distribution but with a different parameterisation.

```
region N NPop
1 11.0 53.0
1  9.0 35.0
1 13.0 56.0
...
```

where `region` is the region number, `N` is the number of individuals detected by the network and `NPop` is the target population counts.

The output for the population count distribution at time instants $t > t_0$ is organized as a list with one element for each time instant $t$. An element for a time instant $t$ is also a list with one or two items, depending on a parameter passed to the function that perform the computations. The first item is a table with descriptive statistics and the second one contains the random values generated according to the corresponding distribution. The structure of these tables is identical with the previous ones.

The third output, the origin-destination matrices for all pairs of time instants *(*`time_from` `time_to`*)* is also a list with one element for each pair of time instants. An element for a pair (`time_from`-`time_to`) is a list with one or two elements, as in the previous case. The first element is a table with descriptive statistics for the origin-destination matrix and the second element of the list gives the random values generated for the population moving from one region to another and it looks like in the example below:

```
region_from region_to iter NPop
1 1 1 40
1 2 1 0
1 3 1 0
...
```

where `NPop` is the random value while `iter` represents the index of the corresponding random value in the whole set.

Below we provide a script that show how to obtain these outputs. The datasets used in this example are provided with the **inference** package and they were obtained using our simulation software. There are three functions providing the main functionalities of this package: *computeInitialPopulation()* for computing population count at $t_0$, *computePopulationT()* for computing population count at $t > t_0$ and *computePopulationOD()* for the Origin-Destination matrices. The actual distribution used for the population counts is specified through the parameter `popDistr`.

```
library(inference, warn.conflicts = FALSE)
path          <- 'extdata'
prefix        <- 'postLocDevice'
postLocPath   <- system.file(path, package = 'inference')
dpFileName    <- system.file(path, 'duplicity.csv', package = 'inference')
rgFileName    <- system.file(path, 'regions.csv', package = 'inference')
omega_r       <- computeDeduplicationFactors(dupFileName = dpFileName,
regsFileName = rgFileName, postLocPrefix = prefix, postLocPath = postLocPath)
nFileName     <- system.file(path, 'nnet.csv', package = 'inference')
nnet          <- readNnetInitial(nFileName)
pRFileName    <- system.file(path, 'pop_reg.csv', package = 'inference')
pRateFileName <- system.file(path, 'pnt_rate.csv', package = 'inference')
grFileName    <- system.file(path, 'grid.csv', package = 'inference')
```

```
params         <- computeDistrParams(omega = omega_r, popRegFileName =
pRFileName, pntRateFileName = pRateFileName, regsFileName = rgFileName,
gridFileName = grFileName)
n_bnb          <- computeInitialPopulation(nnet = nnet, params = params, popDistr
= 'BetaNegBin', rndVal = TRUE)
nnetODFile <- system.file(path, 'nnetOD.zip', package = 'inference')
nt_bnb     <- computePopulationT(nt0 = n_bnb$rnd_values, nnetODFileName =
nnetODFile, rndVal = TRUE)
OD_bnb <- computePopulationOD(nt0 = n_bnb$rnd_values, nnetODFileName =
nnetODFile, rndVal = TRUE)
```

## 3. Computational efficiency and scalability

In this section, we discuss few of the computational efficiency issues and provide our first results of a scalability study.

**destim** has some degree of optimization, as it uses **Rcpp** and **RcppEigen** packages (for sparse linear algebra) in some critical functions and the difference in execution speed between C++ code and pure R code is well-known. We choose to use C++ for some computationally intensive operations (based on level 3 BLAS operations) like LU factorizations or QR decompositions. For other packages in our hierarchy that also work with matrices we used the **Matrix** R package because they only perform simple manipulations of rows or columns of these matrices which run fast enough.

There are some ideas about parallelization and a possible migration to take advantage of some highly optimized linear algebra libraries such as Intel Math Kernel Library (MKL). From now on, we have given priority to the idea of being able to distribute the software fully free with GPL license. At present, the estimation is done individually per device, but efficiency can be substantially improved if it would be done for the whole set of the devices all at once. Other high-performance versions of the BLAS library can be considered too as alternatives to Intel MKL: OpenBLAS (Qian et al., 2013), GotoBLAS or GoToBLAS2 (Goto and van de Geijn, 2008), ATLAS (Whaley et al., 2001; Whaley and Dongarra, 1999).

The default optimization algorithm used to fit the models is *solnp* from package **Rsolnp**. The other possible choice is *constrOptim* from package **stats**. In order to avoid giving a local minimum as solution, there is a parameter that allows the user to execute the optimization several times. The optimizer will be launched with different initial parameters as many times as the number in the argument `retrain`. The model with higher likelihood will be returned.

The most computationally intensive functions of the **deduplication** package (*computeDuplicity*, *computeDuplicityBayesian* and *computeDuplicityTrajectory*) use parallel computations to decrease the execution time. Parallelization is done using the standard techniques found in the **parallel** package: first, the above-mentioned functions build a cluster of working nodes, exports the variables needed for computations to all nodes and then distribute the computations equally among these nodes. While executing the parallel code, all the logical cores of the computer are used. Even using these parallel computations techniques, the execution time could be high, depending on the size of the input data. The most demanding method from the execution time point of view is the *trajectory* method. All the improvements achieved in the geolocation layer will cause an improvement in the efficiency of the Bayesian approaches to execute deduplication. In relation to the *trajectory* approach, as well as in the majority of the processes, it has been taken into account the sparsity of the matrices of distances to speed up the executions. One computationally intensive step of the trajectory method consists in computing the dispersion radius which involves the calculation of the (Euclidean) distances between the rows of a data matrix. For this specific step we use the standard *dist* function but in the future, we can consider replacing it with *rpuDist* from **rpud** package (Yau, 2010) or *gpuDist* from **gputools** package (Buckner et al., 2010) which uses GPU to speed-up the computations. However, the increase in the speed of execution comes with the cost of decreasing the portability of the deduplication package since both **rpud** and **gputools** requires CUDA (Cook, 2012).

The computational intensive functions of the **inference** package (*rNnetEvent*, *netEventOD*) also use parallel computations to decrease the execution time. Parallelization is done using the same standard techniques found in the **parallel** package. The cluster used for parallel computations is a SOCK one under the Windows operating system and a FORK one under Unix-like operating systems.

Functions in **inference** package make use of the processing features of the **data.table** package which implements them very efficiently. Since population at $t$ depends on population at $t - 1$, the computation of the target population distributions at different time instants is inherently sequential. Functions that take a longer time to execute display a progress bar to show how the computations advance.

We plan to test if other parallelization techniques and R packages can provide higher speedups and less memory requirements. To be more specific we intend to test **Rth** package (Matloff and Schmidt, 2015) the provides a function to compute the distance between rows of a matrix and has also parallel implementations for random number generators. Comparing to other GPU-oriented packages mentioned before, **Rth** support several parallel backends: CUDA, OpenMP or Intel TBB. So, if an NVIDIA card is not available to work with CUDA, **Rth** functions uses the multithreading capabilities offered by OpenMP or TBB libraries. Another package that we intend to use in our performance tests is **Rdsm** (Matloff, 2014) which offers a shared-memory approach to parallel processing saving memory and execution time compared with the standard parallel package, although it is available only on Unix-like operating systems.

We run some scalability tests for **deduplication** and **aggregation** packages, computing the execution time and the speedup as functions of the number of threads. The results are presented in figures 2 and 3 which show the execution time and the speedup versus the number of threads. These results emphasize the idea that the two packages show a good scalability.

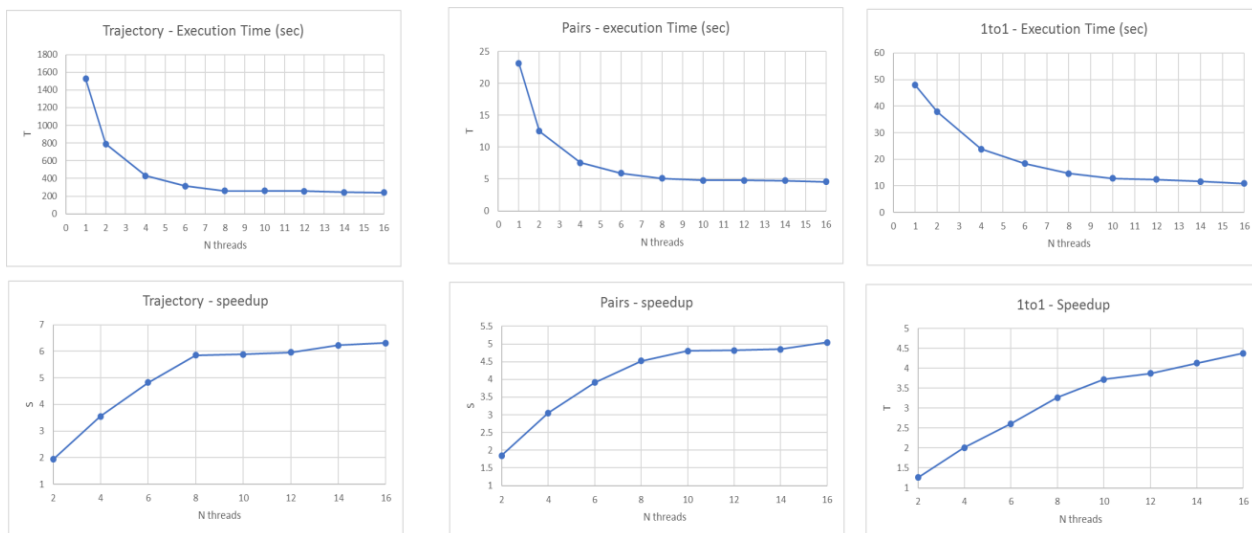

Figure 2. Scalability of the **deduplication** package

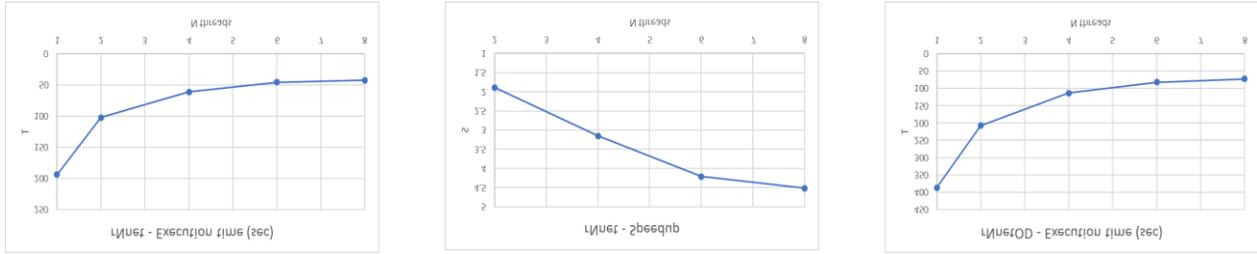
Figure 3. Scalability of the **aggregation** package

## 4. Conclusions and future developments

In this paper we presented a first software implementation of methodological framework developed to integrate the mobile phone data into the production of Official Statistics. This implementation has been done in R which is a free software environment focused on statistical computing. All the code is open, available, and free to be used by any statistician. Following the modularity ideas, an R package has been developed for each module of the process.

Table 1. The R packages

| Module | R package |
|---|---|
| Geolocation | destim |
| Deduplication | deduplication |
| Aggregation | Aggregation |
| Inference | inference |

Regarding the future prospects about the organization of the developed software, we expect to connect all these packages with the aim to make them more user-friendly. The idea is to make another layer over these packages with some high-level functions available to the user. These functions are devoted to obtaining the final results computed in the inference layer which are the goal of the process. They will call functions from the **inference** package which in turn will call functions from the **aggregation** package which will call functions from the **deduplication** package which finally will call functions from the **destim** package. The results will go back to the calling functions until they will reach again the top of the hierarchy and the users will have the answers to their query. The user requests go from the top of the hierarchy to the bottom and the answers go back from the bottom layer to the top. These calls will be optimized (borrowing the principle used by cache memories), for example if a new inference problem implies the same geolocation problem and the results are already available, they will be not be computed again but provided from the storage system. Of course, these optimizations will require some changes of the existing code with a considerable effort but will make all the intermediate steps transparent to users. The current interface will still be available, though.

**Acknowledgments**. This work was partially supported by the European Commission through the European Statistical System [Grant Agreement Number 847375-2018-NL-BIGDATA (ESSnet on Big Data II)].